\documentclass[preprint,nofootinbib,showkeys,longbibliography]{revtex4-1}
\usepackage{epsfig}
\usepackage{amsmath}
\usepackage{amssymb}
\usepackage{hyperref}
\newcommand{\e}{\mathrm{e}}
\newcommand{\h}{\mathrm{h}}
\newcommand{\oo}{\mathrm{o}}

\newcommand{\erf}{\mathrm{erf}}
\newcommand{\erfc}{\mathrm{erfc}}
\newcommand{\op}[1]{\mathsf #1}
\newcommand{\bb}[1]{\mbox{\pmb{$ #1$}}}
\newcommand{\be}{\begin{equation}}
\newcommand{\ee}{\end{equation}}
\newcommand{\ba}{\begin{eqnarray}}
\newcommand{\ea}{\end{eqnarray}}

\newcommand{\lb}{\label}
\begin{document}
\author{Jacek Karwowski}
\affiliation {Institute of Physics, Faculty of Physics, Astronomy and
Informatics,\\
Nicolaus Copernicus University, Grudzi\c{a}dzka 5, 87-100 Toru\'n, Poland\\
email: jka@umk.pl} 
\author{Andreas Savin}
\affiliation {Laboratoire de Chimie Théorique,
CNRS and Sorbonne University, 4 place Jussieu, 
75252 Paris cedex 05, France\\
email: andreas.savin@lct.jussieu.fr}

\title{Erfonium: A Hooke Atom with Soft Interaction Potential}

\begin{abstract}
Properties of {\em erfonium}, a Hooke atom with the Coulomb interaction
potential $1/ \/ r$ replaced by a non-singular $\erf(\mu\,r)/ \/ r$ 
potential are investigated. The structure of the Hooke atom potential and 
properties of its energy spectrum, relative to the ones of the spherical 
harmonic oscillator and of harmonium, are analyzed. It is shown, that 
at a certain value of $\mu$ the system changes its behavior from a 
harmonium-like regime to a harmonic-oscillator-like regime.
  
\end{abstract}
\thanks{This is a preprint of the following chapter: 
Jacek Karwowski and Andreas Savin, A Hooke Atom with Soft Interaction
Potential, published in Recent Progress in Methods and Applications of Quantum
Systems, edited by Ireneusz Grabowski, Karolina S\l{}owik, Erkki J. Br\"andas, 
and Jean Maruani, 2023, Springer Nature, reproduced with permission 
of Springer Nature Switzerland AG Gewerbestrasse 11, 6330 Cham, Switzerland.
The final authenticated version is available online at: 
\url{https://doi.org/10.1007/978-3-031-52078-5_5}.}

\maketitle
\section{Motivation}

The interaction between electrons can be split to two parts:
\be
\lb{01}
\frac{1}{r}=\frac{\erf(\mu\,r)}{r}+\frac{\erfc(\mu\,r)}{r}.
\ee
The first, long-range, term is smooth at $r\,\sim\,0$ and behaves as the 
Coulomb potential at $r\rightarrow\infty$. The second, short-range one,
correctly represents the Coulomb potential at small $r$ and decays
exponentially as $r\,\rightarrow\,\infty$. From this observation stems 
the concept of range separation \cite{GorSav06, Sav11, Sav20},
where the long-range section and the short-range section are treated in 
different ways.

In theories of many-electron systems, the replacement of the singular 
Coulomb potential by a smooth long-range potential, signifficantly 
reduces the computational effort. On the other hand, the short-range 
behavior of the wave function is mostly defined by the universal 
properties of the Coulomb singularity. Therefore the idea of improving 
models which utilize the long-range interactions only, by some
universal corrections describing the short-range properties is very
tempting. Indeed, correcting solutions of the Schr\"odinger equation 
with the long-range part of the interaction potential, by using the
generalized cusp conditions to represent the wave functions in the 
small $r$ area, was recently shown not only feasible, but also very 
accurate \cite{SavKar23}.

The long-range Coulomb interaction potentials, 
\be
\lb{02}
\op{w}(r;\mu)=\frac{\erf(\mu\,r)}{r},
\ee    
have been introduced in several areas of chemistry and physics. For example, 
a model with the Coulomb interaction replaced by $\op{w}(r;\mu)$ correctly 
reproduces some selected properties of the hydrogen atom, harmonium, and 
electron gas \cite{GonAyKarSav16}. In quantum chemistry, the replacement of 
the singular Coulomb potential by a smooth one represented by the error 
function is particularly convenient with the Gaussian basis sets: the 
integrals with this potential are easy to calculate. 
Therefore, the applications of the range separation concept, though most 
common in the density functional theory \cite{Sav11, PerHap21}, extend 
beyond this field \cite{Sav20}.

Also in theory of complex systems, as crystals, liquids, or plasmas, using 
$\erf(\mu\,r)/r$ potential is motivated by its mathematical properties.  
Probably for the first time it was used by Ewald in 1921, well before the 
formulation of quantum mechanics, to replace the Coulomb interaction in order 
to secure the convergence of some series describing infinite systems of 
particles \cite{Ewald921}. A reader interested in these areas of applications 
is referred to a recent paper \cite{DemLev22}, where also references to the
earlier works are supplied.  Due to its convenient mathematical form and
simple relation to the Coulomb potential, $\erf(\mu\,r)/r$ potential is also used 
in many other sections of physics, just to mention so distant fields as
interactions between quarks \cite{LuchaRuppScho992}, or general theory of
relativity \cite{Plam18}.
 
Two electrons with interaction represented by $\op{v}(r)$, in an 
external potential $\op{W}(\bb{r}_1)+\op{W}(\bb{r}_2)$, are described by the 
Hamiltonian
\be
\lb{03}
\op{H}(\bb{r}_1,\bb{r}_2)=-{\textstyle
\frac{1}{2}}\bigtriangleup_{\,\bb{\textstyle{r}}_1}-
{\textstyle \frac{1}{2}}\bigtriangleup_{\,\bb{\textstyle{r}}_2}+
\op{W}(\bb{r}_1)+\op{W}(\bb{r}_2)+\op{v}(r),
\ee
where $r=|\bb{r}|\equiv\left|\bb{r}_1-\bb{r}_2\right|$. Introducing
$\bb{R}=\left(\bb{r}_1+\bb{r}_2\right)/2$ we get 
\be
\lb{04}
\bb{r}_1=\bb{R}+{\textstyle \frac{1}{2}}\,\bb{r},\;\;\;\
\bb{r}_2=\bb{R}-{\textstyle \frac{1}{2}}\,\bb{r}.
\ee 

If $\op{W}(\bb{r}_i)\,\propto\,r_i^2$, $i=1,2$, then the Hamiltonian is 
separable and the two-electron eigenvalue equation splits to two 
spherically-symmetric equations \cite{KestSina62}.  The first one, in
$\bb{R}$, is independent of the interaction potential, and describes the
motion of the center of mass of two electrons in the parabolic confining potential.  
The second equation, in $\bb{r}$, describes the relative motion of two 
interacting electrons in the confining potential $(\omega\,r/2)^2$,
where parameter $\omega$ defines the strength of the confinement. 
We assume that the interaction potential is equal to $\op{w}(r;\mu)$, 
defined in (\ref{02}). The resulting model potential reads
\be
\lb{05}
\op{V}(r)\,\equiv\,\op{V}(r;\omega,\mu)\,=\,\frac{\erf(\mu\,r)}{r}+
\left(\frac{\omega\,r}{2}\right)^2. 
\ee
The elimination of the angular dependence 
from the second equation gives an infinite set of radial equations labelled 
by the angular momentum quantum number $\ell$
\be
\lb{06}
\op{H}(r;\ell,\omega,\mu)\,\phi_{n,\ell}(r)=
E_{n,\ell}(\omega,\mu)\,\phi_{n,\ell}(r),\;\;\;n,\ell=0,1,2,\ldots,
\ee
where $\phi_{n,\ell}(r)=r\,\psi_{n,\ell}(r)$ is the reduced radial function,
$\psi_{n,\ell}(r)$ is the radial part of the one-particle wave function, and
\be
\lb{07}
\op{H}(r;\ell,\omega,\mu)=\left(-\frac{d^2}{dr^2}+\frac{\ell(\ell+1)}{r^2}+
\op{V}(r;\omega,\mu)\right)
\ee
is the radial Hamiltonian.

If the interaction potential $\op{v}(r)$ is repulsive then its energy 
spectrum is continuous. The introduction of a parabolic confinement leads 
to the discretization of the spectrum. Two confined electrons form a bound 
system which resembles a two-electron atom. It is referred to as a Hooke atom 
(due to the confining Hooke force). If the two particles interact by the Coulomb 
potential, the system is called {\em harmonium}. We propose here the name 
{\em erfonium} for a Hooke atom with the Coulomb interatcion 
between electrons replaced by $\op{w}(r;\mu)$. Erfonium is a generalization and 
unification of two well-studied systems: the spherical harmonic oscillator and 
harmonium.  Its radial Hamiltonian (\ref{07}) transforms to the Hamiltonian of 
the spherical harmonic oscillator if $\mu=0$, and to the Hamiltonian of harmonium, 
if $\mu\rightarrow\infty$  (\ref{10}). Accordingly, at these two limits the 
eigenvalues $E_{n,\ell}(\omega,\mu)$, and other quantities characterizing the
system, approach the corresponding quantities of the harmonic oscillator and 
of harmonium. They are marked hereafter by superscripts ${}^\oo$ and ${}^\h$, 
respectively. 

The close relation between the Coulomb and $\op{w}(r;\mu)$ potentials can 
be seen by comparig their integral representations. We have
\be
\lb{08}
\frac{1}{r}=\frac{2}{\sqrt{\pi}}\int_0^\infty\,\e^{-(x\,r)^2}\,dx\;\;\;\;\;\;
\mbox{and}\;\;\;\;\;\;
\frac{\erf(\mu\,r)}{r}=\frac{2}{\sqrt{\pi}}\int_0^\mu\,\e^{-(x\,r)^2}\,dx.
\ee 
Then, 
\be
\lb{09}
\frac{1}{r}\,=\,\lim_{\mu\rightarrow\infty}\,\frac{\erf(\mu\,r)}{r}
\ee
and at the limit of large $\mu$ the model potential (\ref{05}) transforms to the 
potential of harmonium: 
\be
\lb{10}
\op{V}(r;\omega,\mu)\,\mathrel{\raisebox{-6pt}
{$\stackrel{\displaystyle{\rightarrow}}{\scriptstyle{\mu \to \infty}}$}}
\,\op{V}^\h(r;\omega)=\frac{1}{r}+\left(\frac{\omega\,r}{2}\right)^2.
\ee

The $\ell$-dependent term in equation (\ref{07}) describes the 
centrifugal force, and together with the model potential is called the
effective radial potential:
\be
\lb{11}
\mathcal{V}(r;\ell,\omega,\mu)\,=\,\frac{\ell(\ell+1)}{r^2}+
\op{V}(r;\omega,\mu).
\ee
Notice, that the parameter $\mu$ in 
$\mathcal{V}(r;\ell,\omega,\mu)$ describes the adiabatic connection 
between spherical harmonic oscillator
$\left[\mathcal{V}(r;\ell,\omega,0)\right]$, and harmonium 
$\left[\mathcal{V}(r;\ell,\omega,\infty)\right]$.  

For $\ell=0$, $\mathcal{V}(r;\ell,\omega,\mu)$ is equal to the model 
potential (\ref{05}): 
\be
\lb{12}
\mathcal{V}(r;0,\omega,\mu)\,=\,\op{V}(r;\omega,\mu).
\ee
But, for $l>0$ these two potentials are notably different. The most 
significant difference is in the area of small $r$. At $r=0$, 
\be
\lb{13}
\lim_{r\rightarrow\,0}\mathcal{V}(r;0,\omega,\mu)\,=\,
\lim_{r\rightarrow\,0}\op{V}(r;\omega,\mu)=\frac{2\,\mu}{\sqrt{\pi}}, 
\ee
while
\be
\lb{14}
\mathcal{V}(r;\ell,\omega,\mu)\mathrel{\raisebox{-6pt}
{$\stackrel{\displaystyle{\sim}}{\scriptstyle{r \to 0}}$}}
\frac{\ell\,(\ell+1)}{r^2}
\ee
i.e.  it is singular if $\ell\,\ne\,0$.

This paper is aimed at the exploration of some general properties of 
erfonium. General characteristics of the model potential (\ref{05}) are presented
in the next section. The effective radial potential (\ref{11}) is discussed
in section 3. The energy spectrum of erfomium is analyzed in section 4.
Final remarks complete the work. All equations and numerical values are here 
expressed in the Hartree atomic units.  

\section{The model potential}
 
The radial potential of erfonium (\ref{05}) can be expressed as
\ba
\lb{15}
\op{V}(r;\omega,\mu)&=&\mu\,\left[\frac{\erf(\mu\,r)}{(\mu\,r)}
+\frac{2\,q}{3\,\sqrt{\pi}}(\mu\,r)^2\right]\\[5pt] 
\lb{16}
&=&\frac{2\mu}{\sqrt{\pi}}\left[1-\frac{(1-q)}{3\cdot1!}(\mu\,r)^2+
\frac{(\mu\,r)^4}{5\cdot2!}-\frac{(\mu\,r)^6}{7\cdot3!}+\cdots\right]
\ea
where
\be
\lb{17}
q=\frac{3\omega^2\sqrt{\pi}}{8\,\mu^3}.
\ee

Potential $\op{V}(r;\omega,\mu)$ has interesting scaling properties.
According to equation (\ref{15}) 
\be
\lb{18}
\op{V}(r/\mu;\omega,\mu)=\mu\,\left(\frac{\erf(r)}{r}+
\frac{2\,q}{3\,\sqrt{\pi}}\,r^2\right).
\ee
As we can see, all potentials with the same value of $q$ have
exactly the same shape - they only differ by units in which $\op{V}$ and $r$
are expressed. Therefore, for an arbitrary pair $\left(\mu_1,\mu_2\right)$
\be
\lb{19}
\frac{\op{V}(r/\mu_1;\omega_{\,1},\mu_1)}{\op{V}(r/\mu_2;\omega_{\,2},\mu_2)}=
\frac{\mu_1}{\mu_2}
\ee
if
\be
\lb{20}
\omega_{\,2}=\omega_{\,1}\,\left(\frac{\mu_2}{\mu_1}\right)^{3/2},
\ee
i.e. if the cofinement parameters $\omega_{\,1}$ and $\omega_{\,2}$
are related to $\mu_1$ and $\mu_2$ as
\be
\lb{21}
\left(\frac{\omega_{\,1}}{\omega_{\,2}}\right)^2=
\left(\frac{\mu_1}{\mu_2}\right)^3.
\ee
For example, the potential energy curve corresponding to ($\mu=1$,
$\omega=1$) is, except for scaling, the same as the curve for 
($\mu=10^{-2/3}\,\approx\,0.2154$, $\omega\,=\,0.1$) or for 
($\mu=100^{-2/3}\,\approx\,0.0464$, $\omega\,=\,0.01$). 
From equation (\ref{18}) we can get similar scaling relations for the
derivatives of $\op{V}$ with respect to $r$:  
\be
\lb{22}
\frac{\op{V}^{(n)}(r/\mu_1;\omega_{\,1},\mu_1)}
{\op{V}^{(n)}(r/\mu_2;\omega_{\,2},\mu_2)}
=\left(\frac{\mu_1}{\mu_2}\right)^{n+1},
\ee
where $\op{V}^{(n)}$ is the $n$-th derivative of $\op{V}$ with respect to
$r$.

The first and the second derivatives of $\op{V}(r;\omega,\mu)$ are expressed
as
\ba
\lb{23}
\op{V}^\prime(r;\omega,\mu)&=&
\mu^2\,\left[-\frac{\erf(\mu\,r)}{(\mu\,r)^2}+
\frac{2}{\sqrt{\pi}}\frac{\e^{-(\mu\,r)^2}}{(\mu\,r)}+
\frac{4\,q}{3\,\sqrt{\pi}}(\mu\,r)\right]\\[5pt]
\lb{24}
&=&\frac{2\,\mu^2}{\sqrt{\pi}}\left[-\frac{2\,(1-q)}{3\cdot1!}(\mu\,r)+
\frac{4}{5\cdot2!}(\mu\,r)^3-\frac{6}{7\cdot3!}(\mu\,r)^5+\cdots\right];
\ea
\ba
\lb{25}
\op{V}^{\prime\prime}(r;\omega,\mu)&=&
-\frac{2}{r}\,\frac{d\op{V(r;\omega,\mu)}}{dr}-\mu^3\,\frac{4}{\sqrt{\pi}}
\left(\e^{-(\mu\,r)^2}-q\right)\\[5pt]
\lb{26}
&=&\frac{2\,\mu^3}{\sqrt{\pi}}\left[-\frac{2\,(1-q)}{3\cdot1!}+
\frac{3\cdot4}{5\cdot2!}(\mu\,r)^2-
\frac{5\cdot6}{7\cdot3!}(\mu\,r)^4+\cdots\right].
\ea

For finite $\mu$, $\op{V}^\prime(0;\omega,\mu)=0$, and the potential has an 
extremum at $r=0$ -- a minimum if $q>1$ and a maximum if $q<1$ [cf.
equations (\ref{24}) and (\ref{26})]. If $q\ne{1}$ then for small $r$, 
the potential is a quadratic function of $r$. If $q=1$ then, at
$r\,\sim\,0$, the dependence on $r$ is quartic. The threshold value of
$\mu$ corresponding to $q=1$ is equal to
\be
\lb{27}
\mu_0(\omega)=\frac{1}{2}\left(3\sqrt{\pi}\omega^2\right)^{1/3}. 
\ee
Then, $\op{V}(r;\omega,\mu)$ at $r=0$ has a minimum if $\mu<\mu_0(\omega)$ 
and a maximum if $\mu>\mu_0(\omega)$. If $\mu\rightarrow\infty$ then, for 
$r\,\rightarrow\,0$, $\op{V}\,\sim\,1/r$.
Potential $\op{V}(r;\omega,\mu)$ grows up to infinity with increasing
$r$. Therefore, a maximum at $r=0$ implies the existence of a minimum at 
$r=r_\e(\omega,\mu)\equiv r_\e>0$. In the case of harmonium,
\begin{eqnarray}
\lb{28}
r_\e^\h\,\equiv\,r_\e^\h(\omega)&=&\lim_{\mu\rightarrow\infty}r_\e(\omega,\mu)=
\left(\frac{2}{\omega^2}\right)^{1/3},\\
\lb{29}
\op{V}_\e^\h\,\equiv\,\op{V}_\e^\h(\omega)&=&
\lim_{\mu\rightarrow\infty}\op{V}(r_\e^\h;\omega,\mu)\,=\,
\frac{3}{4}\,\left[\omega\,r_\e^\h(\omega)\right]^2\,=
\,\frac{3}{4}\,\left(2\,\omega\right)^{2/3},
\end{eqnarray}
\cite{Taut93,ManMukDie03}. Notice that
\be
\lb{30}
\mu_0(\omega)\,r_\e^\h(\omega)=\left(\frac{3\sqrt{\pi}}{4}\right)^{1/3}
\equiv\alpha
\ee
is independent of $\omega$. 

For a finite $\mu$, the requirement that $\op{V}^\prime(r_\e;\omega,\mu)=0$ 
yields: 
\be
\lb{31}
{\widetilde{r}_\e\,}^3=\erf(\mu\,r_\e)-
\frac{2\,\mu\,r_\e}{\sqrt{\pi}}\,\e^{-(\mu\,r_\e)^2},
\ee
where
\[ 
\widetilde{r}_\e\,\equiv\,\widetilde{r}_\e(\omega,\mu)=
\frac{r_\e(\omega,\mu)}{r_\e^\h(\omega)}=
r_\e(\omega,\mu)\,\left(\frac{\omega^2}{2}\right)^{1/3}
\]
is the normalized coordinate of the minimum.
Notice, that rhs of equation (\ref{31}) is equal to $0$ if $\mu=0$ and
approaches $1$ if $\mu\rightarrow\infty$. Its derivative with respect 
to $x=\mu\,r_\e$ is $(4\,x^2/\sqrt{\pi})\,\e^{-x^2}\,>\,0$. Therefore,
with increasing $\mu$, $r_\e$ monotonically increases from $0$ to $r_\e^\h$. 
The dependence of $\widetilde{r}_\e$ on $\mu$, obtained by numerical solving
equation (\ref{31}), is presented, for several values of omega, in 
the upper left panel of Fig.~1. 

Using equation (\ref{30}) we can rewrite equation (\ref{31}) as
\be
\lb{32}
{\widetilde{r}_\e\,}^3\,=\,\erf(\widetilde{\mu}\,\widetilde{r_\e}\,\alpha)-
\frac{2\,\widetilde{\mu}\,\widetilde{r_\e}\,\alpha}{\sqrt{\pi}}
\,\e^{-(\widetilde{\mu}\,\widetilde{r_\e}\,\alpha)^2},
\ee
where
\[
\widetilde{\mu}\equiv\widetilde{\mu}(\omega)=\frac{\mu}{\mu_0(\omega)}.
\]
Equation (\ref{32}) shows, that the relation between $\widetilde{r}_\e$ and
$\widetilde{\mu}$ is independent of $\omega$. It is plotted in the upper right
panel of Fig.~1. 
\begin{figure}[t]
\lb{f01}
\includegraphics[width=0.90\textwidth]{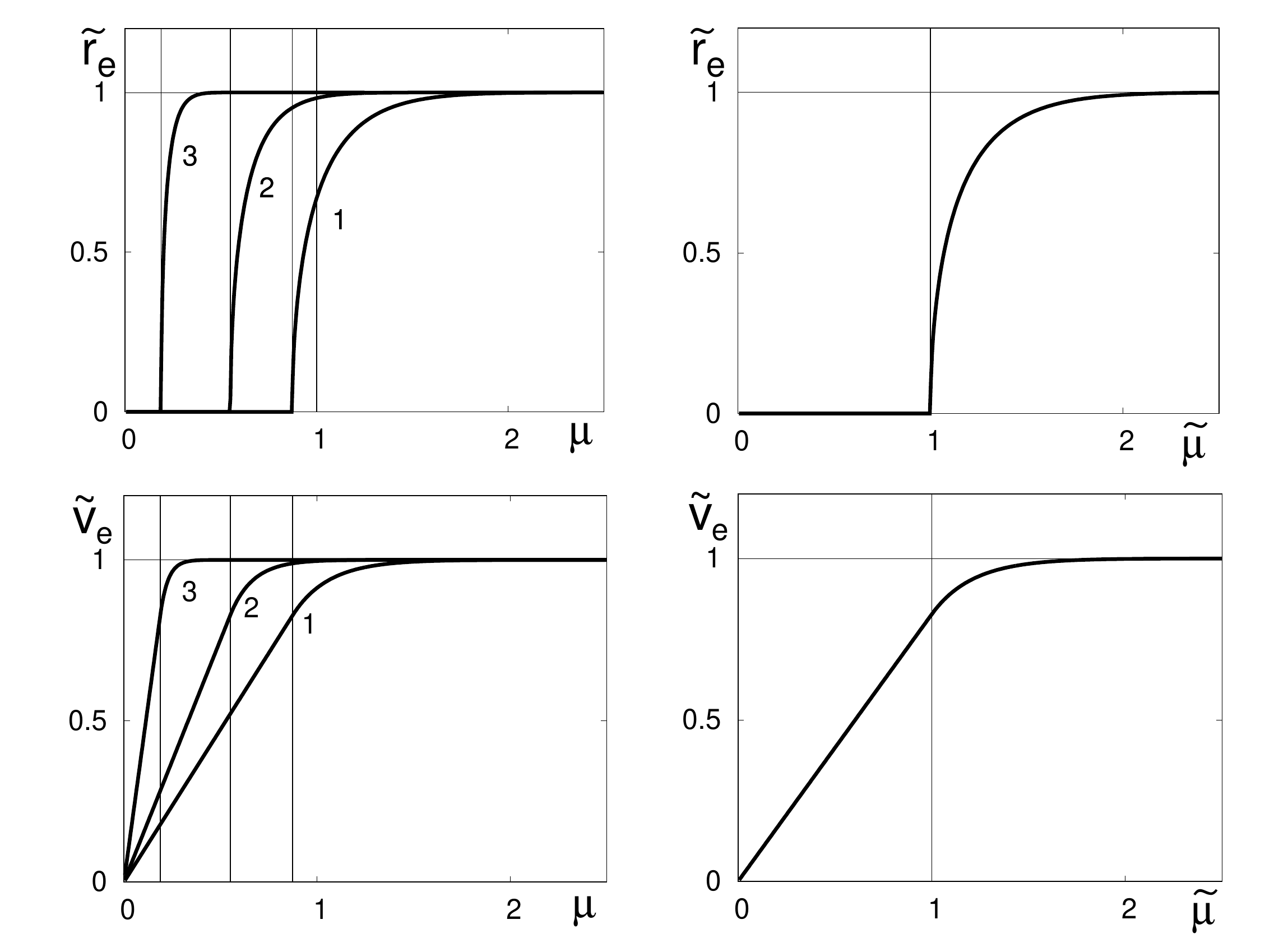}
\caption{Plots of $\widetilde{r}_\e(\omega,\mu)$ (top row)
and $\widetilde{\op{V}}_\e(\omega,\mu)$ (bottom row)
versus $\mu$ (left column) and versus $\widetilde{\mu}$ (right column).  
Curves $1$, $2$, $3$, correspond, respectively, to $\omega=1.0,0.5,0.1$.  
The vertical lines mark $\mu_0(1.0)\approx0.8727$, $\mu_0(0.5)\approx0.5498$, 
and $\mu_0(0.1)\approx0.1880$.  In the right panels the curves overlap: 
$\widetilde{r}_\e(\omega,\widetilde{\mu})$ and 
$\widetilde{\op{V}}_\e(\omega;\widetilde{\mu})$ are the same for all values 
of $\omega$. The parameter $\mu$ is expressed in $\mathrm{bohrs}^{-1}$.} 
\end{figure}      

As one can see consulting equations (\ref{15}) and (\ref{23}), 
\be
\lb{33}
\op{V}_\e\equiv\op{V}_\e(\omega,\mu)\equiv\op{V}(r_\e;\omega,\mu)=
\frac{3}{4}\,(\omega\,r_\e)^2+\frac{2\mu}{\sqrt{\pi}}\,\e^{-(\mu\,r_\e)^2}.
\ee  
The normalized potential at its minimum is defined as
\be
\lb{34}
\widetilde{\op{V}}_\e\equiv\widetilde{\op{V}}_\e(\omega,\mu)=
\frac{\op{V}_\e(\omega,\mu)}{\op{V}^\h(\omega)}.
\ee
Using equations (\ref{29}), (\ref{30}), and (\ref{33}), we get
\be
\lb{35}
\widetilde{\op{V}}_\e={\widetilde{r}_\e\,}^2+\frac{\widetilde{\mu}}{\alpha^2}
\e^{-(\widetilde{\mu}\,\widetilde{r}_\e)^2}.
\ee
Since $\widetilde{r}_\e$, as a function of $\widetilde{\mu}$, is
$\omega$-independent, equation (\ref{35}) implies that also the relation
between $\widetilde{\op{V}}_\e$ and $\widetilde{\mu}$ does not depend on
$\omega$. Plots of $\widetilde{\op{V}}_\e$ versus $\mu$, for three values
of $\omega$ are presented in the lower left panel of Fig.~1.   
The three curves collapse into one in the lower right panel where they are 
plotted against $\widetilde{\mu}(\omega)$. \\

\begin{figure}
\lb{f02}
\includegraphics[width=1.0\textwidth]{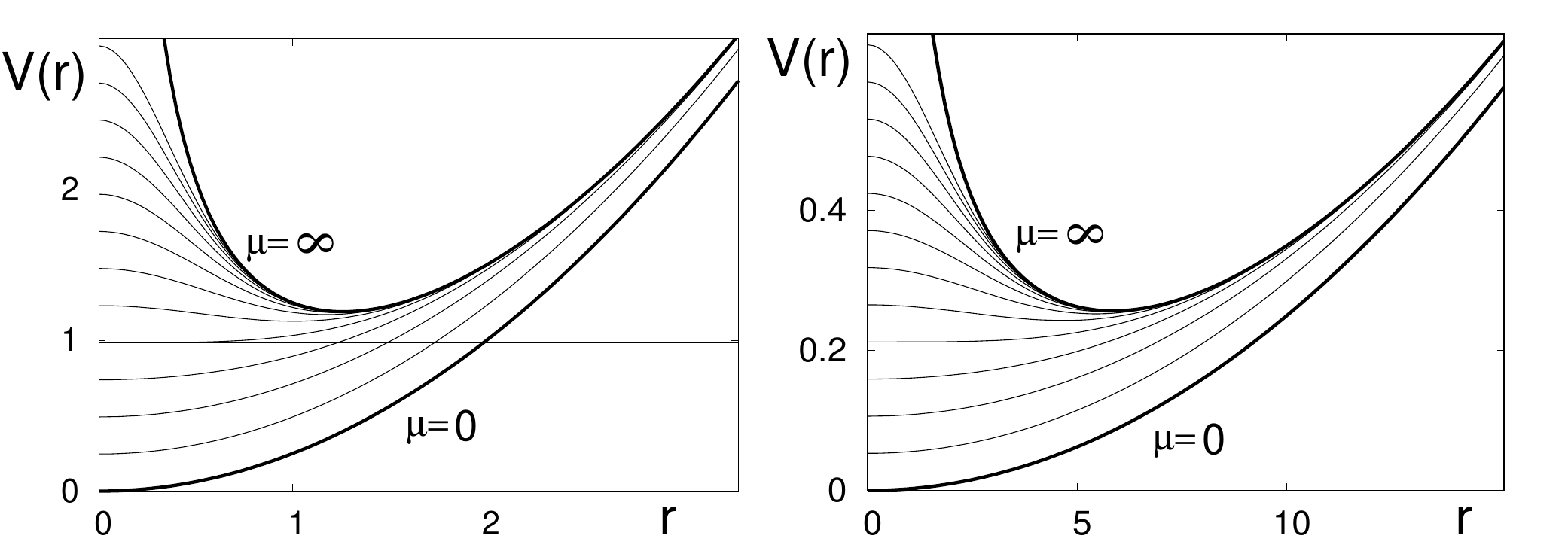}
\caption{Potentials $\op{V}(r)\equiv\op{V}(r;\omega,\mu)$, in 
$\mathrm{hartrees}$, versus $r$, in $\mathrm{bohrs}$. Left panel:  
$\omega=1.0$ [$\mu_0(1.0)=0.8727$]; right panel $\omega=0.1$ 
[$\mu_0(0.1)=0.1880$]. Thick lines: spherical harmonic oscillator ($\mu=0$),
and harmonium ($\mu=\infty$). Thin lines: $\op{V}(r;\omega,\mu_n)$, 
$\mu_n=\mu_0(\omega)\cdot(n/4) $, $n=1,\ldots,12$. The horizontal lines
correspond to $q=1$, i.e. to $\op{V}(0)=2\,\mu_0(\omega)/\sqrt{\pi}$.} 
\end{figure}
\begin{figure}
\lb{f03}
\includegraphics[width=0.99\textwidth]{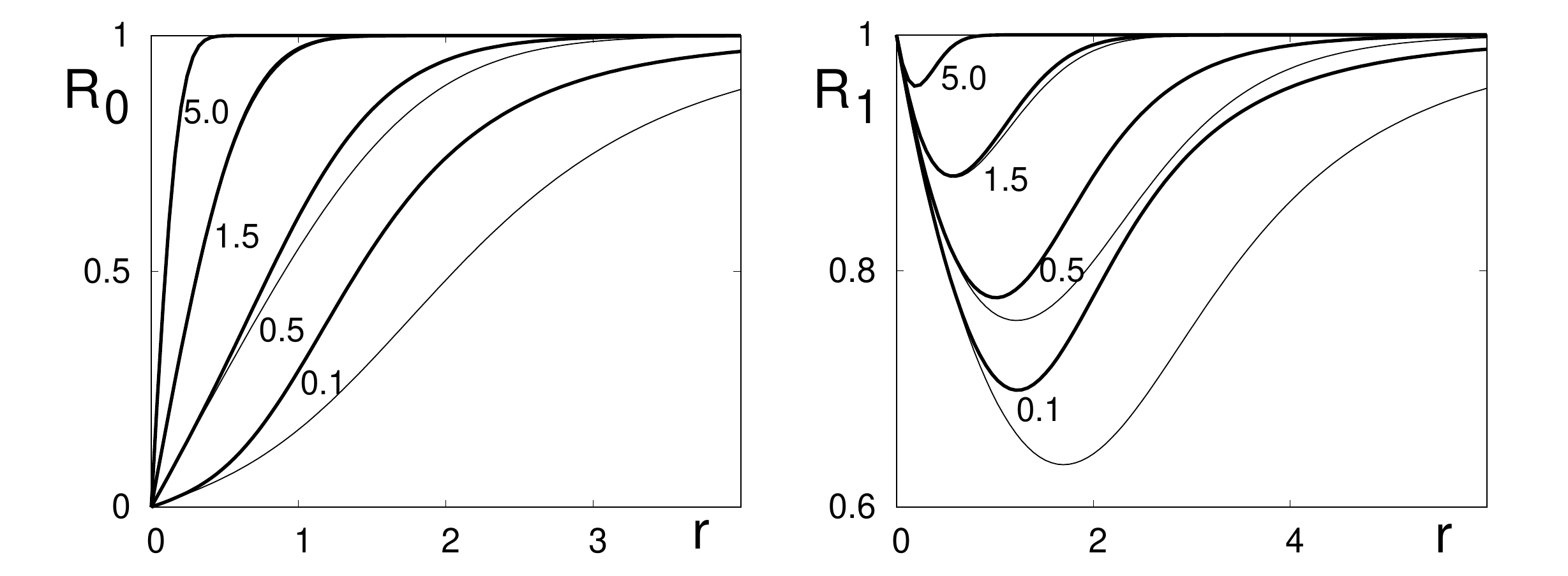}
\caption{Ratio of potentials of erfonium and harmonium versus
$r$ (in $\mathrm{bohrs}$) for $\ell=0$ (left panel) and for $\ell=1$ (right panel). 
Thick lines - $\omega=1.0$; thin lines - $\omega=0.5$. The lines are marked
by the values of $\mu=\,0.1,\,0.5,\,1.5,\,5.0$. If $\mu>1$ then the lines 
corresponding to different $\omega$ overlap.} 
\end{figure}
Plots of $\op{V}(r;\omega,\mu)$ for $\omega=1.0$ and for $\omega=0.1$ are 
shown in Fig.~2. As one can see, the curves correspnding to large $\mu$ are 
close to the potential of harmonium and the ones corresponding to small $\mu$ 
to the potential of the spherical harmonic oscillator.  The shapes of the
left- and of the right-panel potentials ($\omega=1$ and $\omega=0.1$, respectively), 
are the same, but the ranges of units differ by the scaling factor 
$100^{1/3}=4.6416$.

Similarities and differences between erfonium and harmonium potentials
are well reflected by their ratio:
\be
\lb{36}    
R_0(r;\omega,\mu)=\frac{\op{V}(r;\omega,\mu)}{\op{V}^\h(r;\omega)}=
\frac{\erf(\mu\,r)+\omega^2\,r^3/4}{1+\omega^2\,r^3/4}.
\ee
If $\ell=0$ then, for given $\omega$ and $\mu$, the ratio is an increasing 
function of $r$. At the limit of $r\rightarrow\infty$, or $\mu\rightarrow\infty$,
or $\omega\rightarrow\infty$, the ratio is equal to $1$. If $r\,=\,0$,
or $\mu=\omega=0$, then $R_0$\,=\,0. Plots of $R_0(r;\omega,\mu)$ versus $r$, for 
several values of $\omega$ and $\mu$ are shown in the left panel of Fig.~3.
As one can see, the $\omega$-dependence is essential for 
small values of $\mu$ only. One can also notice a fast convergence of 
$\;\op{V}(r;\omega,\mu)$ to $\op{V}^\h(r;\omega)$ with increasing $\mu$. 

\section{The effective potential}

The shape of $\mathcal{V}(r;\ell,\omega,\mu)$, similarly as the shape of 
$\op{V}(r;\omega,\mu)$, is conserved under some specific transformations. 
Equation (\ref{11}) can be rewritten as
\be
\lb{37}
\mathcal{V}(r/\mu;\ell,\omega,\mu)=\mu\left[\frac{\Lambda}{r^2}+
\frac{\erf(r)}{r}+\frac{\omega^2\,r^2}{4\,\mu^3}\right],
\ee
where $\Lambda=\ell\,(\ell+1)\,\mu$. Then,
\be
\lb{38}
\frac{\mathcal{V}(r/\mu_1;\ell_1,\omega_{\,1},\mu_1)}
{\mathcal{V}(r/\mu_2;\ell_2,\omega_{\,2},\mu_2)}=\frac{\mu_1}{\mu_2}
\ee
if 
\be
\lb{39}
\left(\frac{\omega_{\,1}}{\omega_{\,2}}\right)^2=
\left(\frac{\mu_1}{\mu_2}\right)^3,\;\;\;\;
\mbox{and}\;\;\;\;\Lambda_1=\Lambda_2,     
\ee
[compare eq. (\ref{21})]. This means, that two effective potentials have the 
same shape if,
\be
\lb{40}
\mu_2=\mu_1\,\frac{\ell_1(\ell_1+1)}{\ell_2(\ell_2+1)},\;\;\;\;\mbox{and}\;\;\;\;
\omega_2=\omega_1\,\left[\frac{\ell_1(\ell_1+1)}{\ell_2(\ell_2+1)}\right]^{3/2}
\ee

\begin{figure}
\lb{f04}
\includegraphics[width=1.0\textwidth]{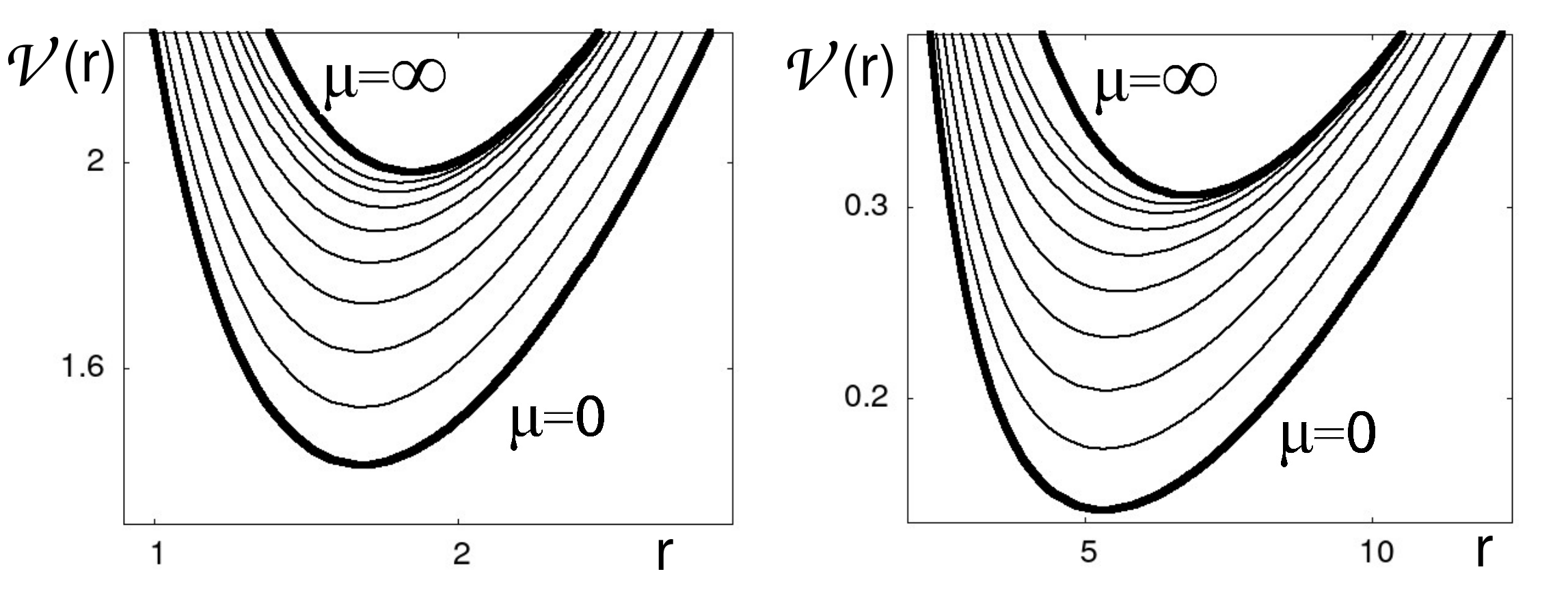}
\caption{The effective potential 
$\mathcal{V}(r)\equiv\mathcal{V}(r;\ell,\omega,\mu)$ (in $\mathrm{hartrees}$) 
for $\ell=1$, as functions of the inter-electronic distance $r$ 
(in $\mathrm{bohrs}$). Left panel - $\omega=1.0$; right panel - $\omega=0.1$. 
The lowest curves represent the effective potentials of the spherical 
harmonic oscillator ($\mu=0$), and the upper ones of harmonium ($\mu=\infty$). 
The curves between these two limits correspond to $\mu=n/10$ (left), 
and $\mu=n/35$ (right), for $n=1,\ldots,8$.}
\end{figure}
As it follows from equations (\ref{10}) and (\ref{11})
\be
\lb{41}
\mathcal{V}(r;\ell,\omega,\mu)\,\mathrel{\raisebox{-6pt}
{$\stackrel{\displaystyle{\rightarrow}}{\scriptstyle{\mu \to \infty}}$}}
\mathcal{V}^\h(r;\ell,\omega)\;\;\;\;\;\mbox{and}\;\;\;\;\;
\mathcal{V}(r;\ell,\omega,\mu)\,\mathrel{\raisebox{-6pt}
{$\stackrel{\displaystyle{\rightarrow}}{\scriptstyle{\mu \to 0}}$}}
\mathcal{V}^\oo(r;\ell,\omega),
\ee
where
\ba
\lb{42}
\mathcal{V}^\h(r;\ell,\omega)&=&\frac{\ell(\ell+1)}{r^2}+\frac{1}{r}
+\left(\frac{\omega\,r}{2}\right)^2,\\
\lb{43}
\mathcal{V}^\oo(r;\ell,\omega)&=&\frac{\ell(\ell+1)}{r^2}
+\left(\frac{\omega\,r}{2}\right)^2,
\ea
are, respectively, the effective potentials of harmonium and of the
spherical harmonic oscillator.
Since $0\,\le\,\op{w}(r,\mu)\,\le\,1/r$, we have
\be
\lb{44}
\mathcal{V}^\oo(r;\ell,\omega)\,\le\,\mathcal{V}(r;\ell,\omega,\mu)\,
\le\,\mathcal{V}^\h(r;\ell,\omega).
\ee
The general shapes of $\mathcal{V}^\oo(r;\ell,\omega)$ and 
$\mathcal{V}^\h(r;\ell,\omega)$ are similar: regular potential wells
approaching infinity in the same way if $r\rightarrow{0}$ and 
$r\rightarrow\infty$. Because $\op{w}(r,\mu)$ is regular and monotonic,
the shapes of $\mathcal{V}(r;\ell,\omega,\mu)$ are similar.
In Fig.~4 plots of $\mathcal{V}(r;\ell,\omega,\mu)$ for $\ell=1$, 
$\omega=1$ (left panel), $\ell=1$, $\omega=0.1$ (right panel), and 
for several values of $\mu$ are presented.  
The ranges of $r$ and $\mathcal{V}$ in the two panels are selected so 
that the shapes of the curves are similar [the shapes cannot be the same,
because $\ell_1=\ell_2=1$, cf. eq. (\ref{39})].   

\begin{figure}
\lb{f05}
\includegraphics[width=0.95\textwidth]{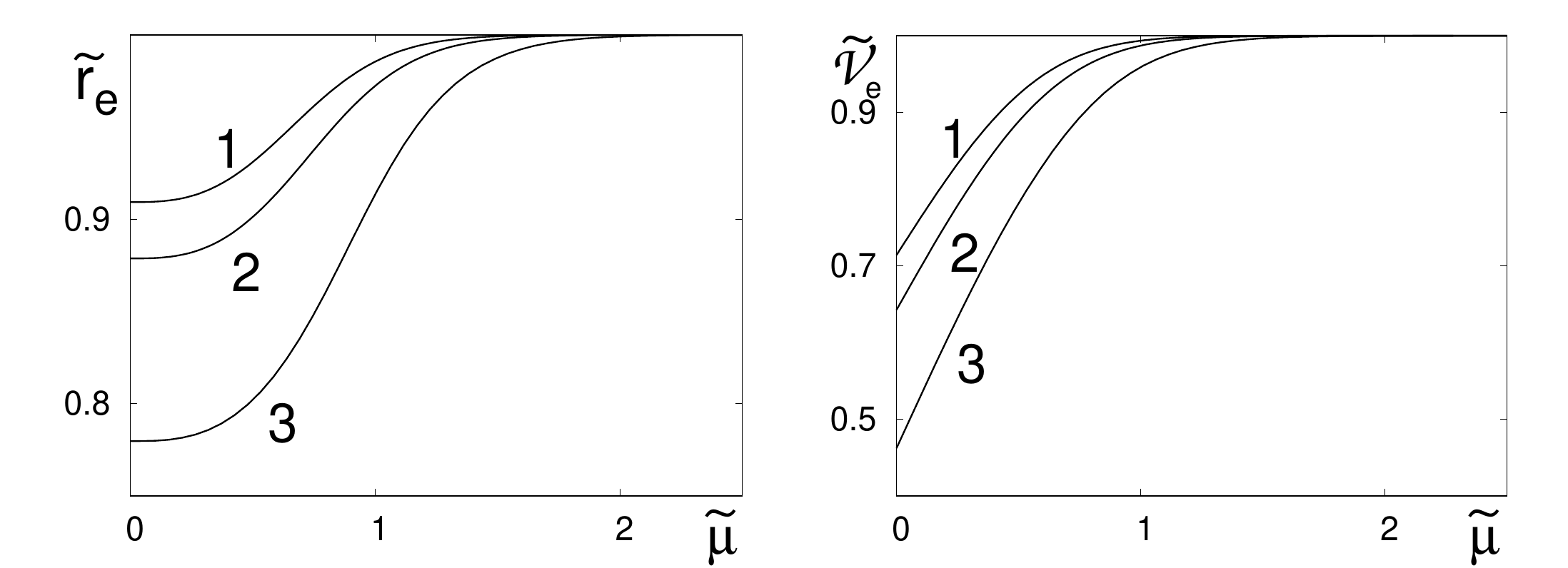}
\caption{Plots of the normalized coordinates of the minimum of the effective
potential for $\ell=1$ and $\omega=1.0,0.5,0.1$ -- curves $1$, $2$, $3$, 
respectively, as functions of $\widetilde{\mu}$. In the left panel
$\widetilde{r}_\e\,\equiv\,\widetilde{r}_\e(\ell,\omega,\mu)$; in the 
right one $\widetilde{\mathcal{V}}_\e\,\equiv\,\widetilde{\mathcal{V}}_\e
(\ell,\omega,\mu)$.} 
\end{figure}  
If $\ell>0$ then the effective potential $\mathcal{V}(r;\ell,\omega,\mu)$ 
has only one extremum: a minimum, at $r=r_\e(\ell,\omega,\mu)$. 
The values of $r_\e$ can be obtained from the condition
\[
\left.\frac{\mathcal{V}(r;\ell,\omega,\mu)}{dr}\right|_{r=r_\e}\,=\,0,
\]
transformed to 
\be
\lb{45}
(r_\e)^4=\frac{2}{\omega^2}\left[2\ell(\ell+1)+\mu\,r_\e^2
\,\left(\frac{\erf(\mu\,r_\e)}{\mu\,r_\e}-\frac{2}{\sqrt{\pi}}\,
\e^{-(\mu\,r_\e)^2}\right)\right],
\ee
where $r_\e$ stands here for $r_\e(\ell,\omega,\mu$). Equation (\ref{45}) 
has to be solved numerically, but one can see that 
$r_\e(\ell,\omega,\mu_1)\,\ge\,r_\e(\ell,\omega,\mu_2$), if 
$\mu_1\,>\,\mu_2$. At the limit of $\mu=0$
(harmonic oscillator), 
\be
\lb{46}
r^\oo_\e(\ell,\omega)^4=\frac{4\ell(\ell+1)}{\omega^2}.
\ee 
If $\mu\rightarrow\infty$ (harmonium), then $r_\e^\h$ is a root of a
fourth-order polynomial:
\be
\lb{47}
r^\h_\e(\ell,\omega)^4=\frac{2}{\omega^2}\left[
2\ell(\ell+1)+r^\h_\e(\ell,\omega)\right].
\ee
The explicit expression is given in \cite{ManMukDie03}. 
The second coordinate of this minimum is 
$\mathcal{V}_\e\,\equiv\,\mathcal{V}(r_\e;\ell,\omega,\mu)$, 
The normalized coordinates of the minimum,
\be
\lb{48}
\widetilde{r}_\e\,\equiv\,\widetilde{r}_\e(\ell,\omega,\mu)
=\frac{r_\e(\ell,\omega,\mu)}{r_\e^\h(\ell,\omega)},\;\;\;\;\;
\mbox{and}\;\;\;\;\;
\widetilde{\mathcal{V}}_\e\,\equiv\,
\widetilde{\mathcal{V}}_\e(\ell,\omega,\mu)=
\frac{\mathcal{V}(r_\e;\ell,\omega,\mu)}{\mathcal{V}^\h(r_\e;\ell,\omega)},
\ee
are plotted versus $\widetilde{\mu}$ in, respectively, left and right 
panels of Fig.~5. Similarly as in the 
case of the model potential (cf. Fig.~1) the minimum of the 
effective potential is approximately located at its asymptotic
position already for $\mu\,\approx\,2\,\mu_0(\omega)$. But, unlike the case 
of $\ell=0$, here both $\widetilde{r}_\e(\ell,\omega,\widetilde{\mu})$ 
and $\widetilde{\mathcal{V}}_\e(\ell,\omega,\widetilde{\mu})$ depend 
on $\omega$. 

Finally, the ratio
\be
\lb{49}
R_\ell(r;\ell,\omega,\mu)=
\frac{\mathcal{V}(r;\ell,\omega,\mu)}{\mathcal{V}^\h(\ell,r;\omega)},
\ee
for $\ell=1$, and for several values of $\omega$ and $\mu$ is plotted
in the right panel of Fig.~3. As in the case of 
$\ell=0$, the $\omega$-dependence is noticeable for small values of 
$\mu$ only. The differences between the effective potentials for finite 
values of $\mu$, and their asymptotic forms is, in the case of $\ell>0$, 
much smaller than for $\ell=0$. This can be seen in Fig.~3: the 
maximum difference between $1$ and $R_1$ is about three times smaller 
than between 1 and $R_0$. 

\section{Spectrum}

\begin{figure}[t]
\lb{f06}
\includegraphics[width=1.0\textwidth]{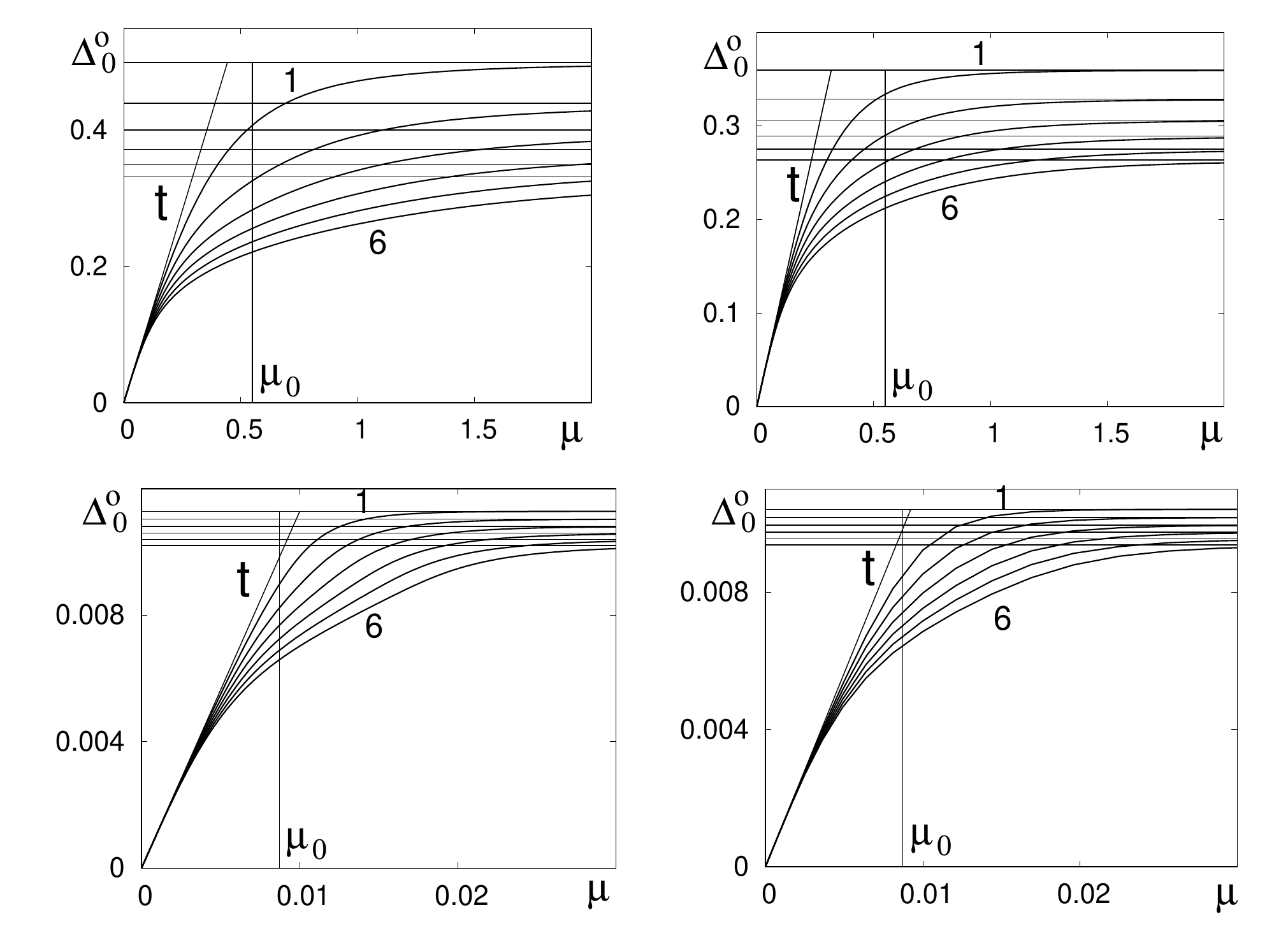}
\caption{Plots of $\Delta_0^\oo$ [eq. (\ref{53})] expressed in
$\mathrm{hartrees}$ versus $\mu$ (in $\mathrm{bohrs^{-1}}$) for 
$\omega=0.5$ (top panels) and $\omega=0.001$ 
(bottom panels); $\ell=0$ (left panels) and $\ell=1$ (right panels). 
The vertical lines mark $\mu_0(\omega)$; the tangent lines $t=t(\mu)$ in the 
bottom-left corners of the panels and the horizontal lines in the upper parts 
of the panels show the asymptotic behavior of  $\Delta_0^\oo(\mu)$ for 
$\mu\rightarrow{0}$ and $\mu\rightarrow\infty$, respectively. The upper 
curves (marked $1$) correspond to the ground states and the lowest ones 
(marked $6$) - to the fifth excited states.}
\end{figure}
At $\mu=0$ the energy spectrum is given by the well known analytic formula
\be
\lb{50}
E_{n,\ell}(\omega,0)\,=\,E_{n,\ell}^\oo(\omega)\,=\,
\omega\,\left(2\,n+\ell+\frac{3}{2}\right).
\ee
According to the Hellmann-Feynman theorem, 
\be
\lb{51}
\frac{\partial\,E_{n,\ell}(\omega,\mu)}{\partial\,\mu}\,=\,
\frac{2}{\sqrt{\pi}}\langle\phi_{n,\ell}\left|\e^{-(\mu\,r)^2}\right|
\phi_{n,\ell}\rangle\,\longrightarrow\,\left\{
\begin{array}{cl}
2/\sqrt{\pi},\;\;\mbox{if}\;\;\mu=0,\\
0,\;\;\mbox{if}\;\;\mu\rightarrow\infty.
\end{array}\right.
\ee
Therefore, for a fixed $\omega$, and $\mu\,\ll\,1$, 
\be
\lb{52}
E_{n,\ell}(\omega,\mu)\,\;\mathrel{\raisebox{-6pt}
{$\stackrel{\displaystyle{\rightarrow}}{\scriptstyle{\mu \to 0}}$}}
\,\;\frac{2\mu}{\sqrt{\pi}}+E_{n,\ell}(\omega,0)\,=\,
\frac{2\mu}{\sqrt{\pi}}+\omega\,\left(2\,n+\ell+\frac{3}{2}\right).
\ee
Differences
\be
\lb{53}
\Delta_0^\oo\,\equiv\,\Delta_0^\oo(n,\ell;\omega,\mu)=E_{n,\ell}(\omega,\mu)-
E^\oo_{n,\ell}(\omega),
\ee
for $n=0,1,\ldots,6$, $\ell=0,1$, and for two values of $\omega$   
differing by factor $500$ ($\omega=0.5$, and $\omega=0.001$) are
plotted, as functions of $\mu$, in Fig.~6. Though the scaling invariance
discussed in previous sections is not valid exactly, one can see that the 
spectra for different values of the confinement parameter have similar
structure.

The average distance between electrons, $\langle\,r\,\rangle$, in the case of 
the harmonic oscillator, is proportional to $1/\sqrt{\omega}$. 
This kind of dependence is approximately valid for a broad range of $\omega$. 
The increasing repulsion between electrons (increasing $\mu$) leads 
to an increase of $\langle\,r\,\rangle$. In the case of $\omega=0.001$ the 
average distance between electrons is of the order of $100\;\mathrm{bohrs}$, 
increasing from about $50\;\mathrm{bohrs}$ at the limit of the harmonic 
oscillator to $130\;\mathrm{bohrs}$ at the limit of harmonium. In such cases 
the influence of the centrifugal term, proportional to $1/r^2$, is  
small -- for small $r$ the wave function is close to $0$. 
This can be seen by comparing the left and the right 
panels in Fig.~6: a difference is visible if $\omega=0.5$, but  
if $\omega=0.001$, then the plots for $\ell=0$ and for $\ell=1$ are nearly
the same -- the relative difference between the ground state energies
of $\ell=0$ and $\ell=1$ is, in the case of $\omega=0.001$, about $1\%$.
The range of the validity of the linear approximation (\ref{52}) 
can be estimated for small $\mu$ by comparing the tangent lines
\be
\lb{54} 
t(\mu)\,=\,\frac{2\,\mu}{\sqrt{\pi}}
\ee
with the curves representing the exact energies.

\begin{figure}
\lb{f07}
\includegraphics[width=1.0\textwidth]{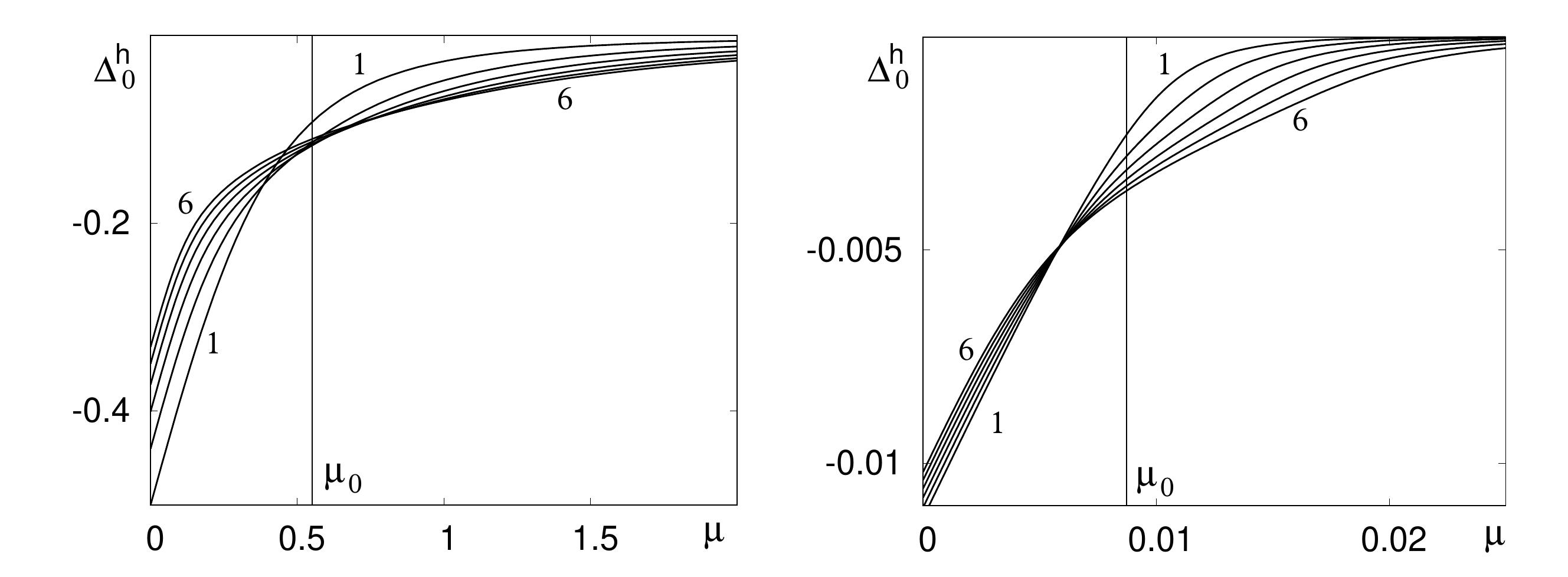}
\caption{Plots of $\Delta_0^\h$ [eq. (\ref{55})] expressed in 
$\mathrm{hartrees}$ versus $\mu$ (in $\mathrm{bohrs}^{-1}$) for $\ell=0$. 
Left panel: $\omega=0.5$; right panel: $\omega=0.001$. 
The vertical lines mark $\mu_0(\omega)$. The curves (marked $1$) correspond to 
the ground state and the ones marked $6$ - to the fifth excited state.}
\end{figure}
The energy of harmonium can be expressed analytically only in some 
special cases, but the structure of its spectrum is well known 
(see e.g. \cite{KarCyr03}). As a function of $\omega$, the spectrum 
can be divided to three parts: 
(1) $\omega\,>\,1$ -- strong confinement dominates the Coulomb interaction,
the electron correlation is weak and we have the harmonic oscillator regime; 
(2) $\omega\,\in\,(10^{-5}\,,\,1)$ -- the 
intermediate regime, the relation between the confinement and the Coulomb 
interaction is, in a sense, similar as in atoms; (3) $\omega\,<\,10^{-5}$ -- 
very weak confinement, the electrons are far apart and their motion is 
strongly correlated. Differences
\be
\lb{55}
\Delta_0^\h\,\equiv\,
\Delta_0^\h(n,\ell;\omega,\mu)=
E_{n,\ell}(\omega,\mu)-E^\h_{n,\ell}(\omega),
\ee
for $n=0,1,\ldots,6$, $\ell=0$, and $\omega=0.5,\,0.001$ are
plotted, as functions of $\mu$, in Fig.~7. Similarly as in the diagrams
presented in Fig.~6, also here one can see a qualitative change of the 
spectrum when $\mu\,\approx\,\mu_0(\omega)$. In the case of the 
harmonic oscillator the distances between the neighboring energy
levels, for given $\ell$ and $\omega$, are constant (independent of $n$). 
In the case of harmonium, these distances increase with increasing $n$ 
\cite{KarCyr03}. When $\mu$ changes from the large to the small values, 
the distances between the energy levels gradually change from the mode of
harmonium to the mode of harmonic oscillator. This can be seen in Fig.~7 --
in the area $\mu\,\approx\,\mu_0(\omega)$, the order of the values of $\Delta_0^\h$
in terms of $n$, changes to the reverse. 
    
\begin{figure}
\lb{f08}
\includegraphics[width=1.0\textwidth]{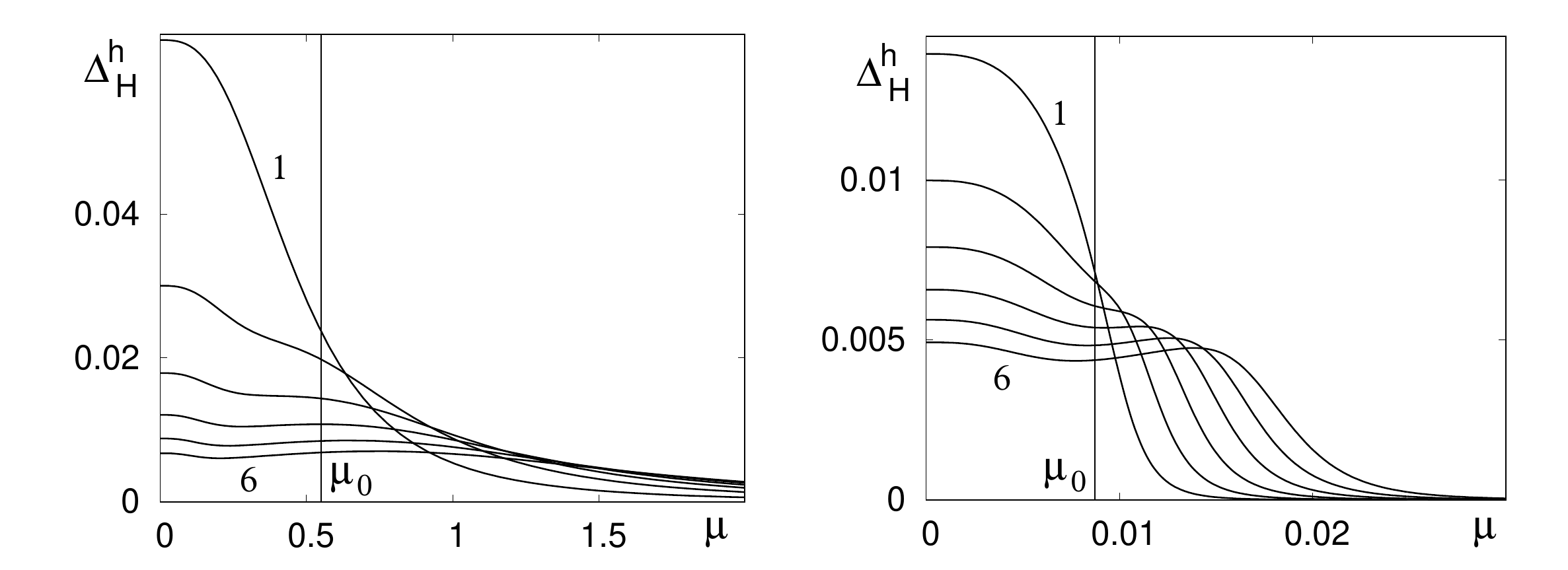}
\caption{Plots of $\Delta_\mathrm{H}^\h\,\equiv\,
\Delta_\mathrm{H}^\h(n,\ell;\omega,\mu)$ [eq. (\ref{56})] expressed in 
$\mathrm{hartrees}$ versus $\mu$ (in $\mathrm{bohrs}^{-1}$), for $\omega=0.5$ 
(left panel) and $\omega=0.001$ (right panel), and $\ell=0$. The vertical 
lines mark $\mu_0(\omega)$. The curves marked $1$ correspond to the ground 
state and the ones marked $6$ - to the fifth excited state.}
\end{figure}
As it was already mentioned, using the smooth long-range interaction
potential instead of the singular Coulomb one, reduces the computational 
effort in the approximate solving the Schr\"odinger equation for
many-electron systems \cite{GorSav06, Sav11, Sav20}. Therefore, one of the 
objectives of studying properties of erfonium, is an exploration
of the possibility of an extrapolation of the energy derived from a model 
with a finite $\mu$ to the limit of electrons interacting by the Coulomb 
force, i.e. the transition from the spectrum of a finite-$\mu$ erfonium to
the spectrum of harmonium. The first step in such a procedure is the 
approximation of the energies of harmonium by the expectation values of 
the harmonium Hamiltonian calculated using the wave functions of erfonium 
\cite{SavKar23}. The resulting correction to the energy, equal to
\be
\lb{56}
\Delta_\mathrm{H}^\h\,\equiv\,\Delta_\mathrm{H}^\h(n,\ell;\omega,\mu)=
\langle\,\phi_{n,\ell}(\mu)|\op{H}^\h|\phi_{n,\ell}(\mu)\,\rangle-
E_{n,\ell}^\h(\omega),
\ee
where
\[
\langle\,\phi_{n,\ell}(\mu)|\op{H}^\h|\phi_{n,\ell}(\mu)\,\rangle\,
\equiv\,\langle\,\phi_{n,\ell}(r;\omega,\mu)|\op{H}(r;\ell,\omega,\infty)|
\phi_{n,\ell}(r;\omega,\mu)\,\rangle,
\] 
is plotted in Fig.~8. As one can see by comparing with Fig.~7, the 
difference between $\left|\Delta_0^\h\right|$ and 
$\left|\Delta_\mathrm{H}^\h\right|$ depends on the value of $\omega$.
If $\omega$ is large (left panels), then 
$\left|\Delta_\mathrm{H}^\h\right|\ll\left|\Delta_0^\h\right|$ (for
$\omega=0.5$ by factor $\approx\,10$). But if $\omega$ is small (right
panels), then both differences are of the same order of magnitude. 

The difference between $E_{n,\ell}(\omega,\mu)$ and
$E_{n,\ell}^\h(\omega)$ can be dramatically reduced if the generalized
cusp conditions are used to represent the wave function at small $r$. 
But, if $\mu$ is smaller than a certain threshold, a further reduction of
the difference using this method proved to be impossible \cite{SavKar23}.  
From the perspective of the present analysis it is understandable that
the threshold value is close to $\mu_0(\omega)$, where the character of the
potential (and of the spectrum) changes from a harmonium-like to an
oscillator-like.

\section{Final remarks}

Erfonium, a Hooke atom with a soft interaction potential 
represented by $\erf(\mu\,r)/r$, makes a bridge between harmonium  
($\mu\,\rightarrow\,\infty$) and the spherical harmonic oscillator ($\mu=0$). 
The parameter $\mu$ establishes an adiabatic connection between these
two limits. An interplay between the effects specific for harmonium and
for the harmonic oscillator creates a rich and interesting pattern which 
can be observed in the structure of the potential and of the spectrum. 
The parameter $\mu_0(\omega)$ determined by the strength of the confinement
(\ref{27}), marks a fuzzy boundary between these two "spheres of influence".
If $\mu>\mu_0(\omega)$ then the regime of harmonium dominates, if
$\mu<\mu_0(\omega)$, then  the features of the harmonic oscillator prevail. 

A vast field for further studies remains open. Particularly interesting (and
difficult) are issues related to the behavior of erfonium at the limit 
of $\omega\,\rightarrow\,0$. Notice that at this limit the spectrum abruptly
changes from entirely discrete, to entirely continuous. If simultaneously
$\mu\,\rightarrow\,0$, then at the limit we get two unconfined and
non-interacting 'electrons'. 

Last, but not least, a broad class of semi-analytic (cf.
\cite{Taut93,ManMukDie03}) and variational approaches may result in precise
analytic approximations of the wave functions and of the energies of erfonium. 
\\

\noindent
{\large \textbf{Acknowledgement}}\\

We thank Prof. Henryk A. Witek (National Chiao Tung University, Hsinchu,
Taiwan) for his constructive remarks and for useful discussions.

\section*{ORCID}
\noindent
Jacek Karwowski: https://orcid.org/0000-0003-1508-2929\\
Andreas Savin: https://orcid.org/0000-0001-8401-8037

\bibliography{m-ir}

\begin{thebibliography}{14}%
\makeatletter
\providecommand \@ifxundefined [1]{%
 \@ifx{#1\undefined}
}%
\providecommand \@ifnum [1]{%
 \ifnum #1\expandafter \@firstoftwo
 \else \expandafter \@secondoftwo
 \fi
}%
\providecommand \@ifx [1]{%
 \ifx #1\expandafter \@firstoftwo
 \else \expandafter \@secondoftwo
 \fi
}%
\providecommand \natexlab [1]{#1}%
\providecommand \enquote  [1]{``#1''}%
\providecommand \bibnamefont  [1]{#1}%
\providecommand \bibfnamefont [1]{#1}%
\providecommand \citenamefont [1]{#1}%
\providecommand \href@noop [0]{\@secondoftwo}%
\providecommand \href [0]{\begingroup \@sanitize@url \@href}%
\providecommand \@href[1]{\@@startlink{#1}\@@href}%
\providecommand \@@href[1]{\endgroup#1\@@endlink}%
\providecommand \@sanitize@url [0]{\catcode `\\12\catcode `\$12\catcode
  `\&12\catcode `\#12\catcode `\^12\catcode `\_12\catcode `\%12\relax}%
\providecommand \@@startlink[1]{}%
\providecommand \@@endlink[0]{}%
\providecommand \url  [0]{\begingroup\@sanitize@url \@url }%
\providecommand \@url [1]{\endgroup\@href {#1}{\urlprefix }}%
\providecommand \urlprefix  [0]{URL }%
\providecommand \Eprint [0]{\href }%
\providecommand \doibase [0]{http://dx.doi.org/}%
\providecommand \selectlanguage [0]{\@gobble}%
\providecommand \bibinfo  [0]{\@secondoftwo}%
\providecommand \bibfield  [0]{\@secondoftwo}%
\providecommand \translation [1]{[#1]}%
\providecommand \BibitemOpen [0]{}%
\providecommand \bibitemStop [0]{}%
\providecommand \bibitemNoStop [0]{.\EOS\space}%
\providecommand \EOS [0]{\spacefactor3000\relax}%
\providecommand \BibitemShut  [1]{\csname bibitem#1\endcsname}%
\let\auto@bib@innerbib\@empty
\bibitem [{\citenamefont {Gori-Giorgi}\ and\ \citenamefont
  {Savin}(2006)}]{GorSav06}%
  \BibitemOpen
  \bibfield  {author} {\bibinfo {author} {\bibfnamefont {P.}~\bibnamefont
  {Gori-Giorgi}}\ and\ \bibinfo {author} {\bibfnamefont {A.}~\bibnamefont
  {Savin}},\ }\bibfield  {title} {\enquote {\bibinfo {title} {Properties of
  short-range and long-range correlation energy density functionals from
  electron-electron coalescence},}\ }\href {\doibase
  10.1103/PhysRevA.73.032506} {\bibfield  {journal} {\bibinfo  {journal} {Phys.
  Rev. A}\ }\textbf {\bibinfo {volume} {73}},\ \bibinfo {pages} {032506}
  (\bibinfo {year} {2006})}\BibitemShut {NoStop}%
\bibitem [{\citenamefont {Savin}(2011)}]{Sav11}%
  \BibitemOpen
  \bibfield  {author} {\bibinfo {author} {\bibfnamefont {A.}~\bibnamefont
  {Savin}},\ }\bibfield  {title} {\enquote {\bibinfo {title} {Correcting model
  energies by numerically integrating along an adiabatic connection and a link
  to density functional approximations},}\ }\href {\doibase 10.1063/1.3592782}
  {\bibfield  {journal} {\bibinfo  {journal} {J. Chem. Phys.}\ }\textbf
  {\bibinfo {volume} {134}},\ \bibinfo {pages} {214108} (\bibinfo {year}
  {2011})}\BibitemShut {NoStop}%
\bibitem [{\citenamefont {Savin}(2020)}]{Sav20}%
  \BibitemOpen
  \bibfield  {author} {\bibinfo {author} {\bibfnamefont {A.}~\bibnamefont
  {Savin}},\ }\bibfield  {title} {\enquote {\bibinfo {title} {Models and
  corrections: {Range} separation for electronic interaction -- {Lessons} from
  density functional theory},}\ }\href {\doibase 10.1063/5.0028060} {\bibfield
  {journal} {\bibinfo  {journal} {J. Chem. Phys.}\ }\textbf {\bibinfo {volume}
  {153}},\ \bibinfo {pages} {160901} (\bibinfo {year} {2020})}\BibitemShut
  {NoStop}%
\bibitem [{\citenamefont {Savin}\ and\ \citenamefont
  {Karwowski}(2023)}]{SavKar23}%
  \BibitemOpen
  \bibfield  {author} {\bibinfo {author} {\bibfnamefont {A.}~\bibnamefont
  {Savin}}\ and\ \bibinfo {author} {\bibfnamefont {J.}~\bibnamefont
  {Karwowski}},\ }\bibfield  {title} {\enquote {\bibinfo {title} {Correcting
  models with long-range electron interaction using generalized cusp
  conditions},}\ }\href {\doibase 10.1021/acs.jpca.2c08426} {\bibfield
  {journal} {\bibinfo  {journal} {J. Phys. Chem. A}\ }\textbf {\bibinfo
  {volume} {127}},\ \bibinfo {pages} {1377--1385} (\bibinfo {year}
  {2023})}\BibitemShut {NoStop}%
\bibitem [{\citenamefont {González-Espinoza}\ \emph
  {et~al.}(2016)\citenamefont {González-Espinoza}, \citenamefont {Ayers},
  \citenamefont {Karwowski},\ and\ \citenamefont {Savin}}]{GonAyKarSav16}%
  \BibitemOpen
  \bibfield  {author} {\bibinfo {author} {\bibfnamefont {C.~E.}\ \bibnamefont
  {González-Espinoza}}, \bibinfo {author} {\bibfnamefont {P.~W.}\ \bibnamefont
  {Ayers}}, \bibinfo {author} {\bibfnamefont {J.}~\bibnamefont {Karwowski}}, \
  and\ \bibinfo {author} {\bibfnamefont {A.}~\bibnamefont {Savin}},\ }\bibfield
   {title} {\enquote {\bibinfo {title} {Smooth models for the {Coulomb}
  potential},}\ }\href {\doibase 10.1007/s00214-016-2007-5} {\bibfield
  {journal} {\bibinfo  {journal} {Theor. Chem. Acc.}\ }\textbf {\bibinfo
  {volume} {135}},\ \bibinfo {pages} {256} (\bibinfo {year}
  {2016})}\BibitemShut {NoStop}%
\bibitem [{\citenamefont {Pernal}\ and\ \citenamefont
  {Hapka}(2022)}]{PerHap21}%
  \BibitemOpen
  \bibfield  {author} {\bibinfo {author} {\bibfnamefont {K.}~\bibnamefont
  {Pernal}}\ and\ \bibinfo {author} {\bibfnamefont {M.}~\bibnamefont {Hapka}},\
  }\bibfield  {title} {\enquote {\bibinfo {title} {Range-separated
  multiconfigurational density functional theory methods},}\ }\href {\doibase
  10.1002/wcms.1566} {\bibfield  {journal} {\bibinfo  {journal} {WIREs Comput.
  Mol. Sci.}\ }\textbf {\bibinfo {volume} {12}},\ \bibinfo {pages} {e1566}
  (\bibinfo {year} {2022})}\BibitemShut {NoStop}%
\bibitem [{\citenamefont {Ewald}(1921)}]{Ewald921}%
  \BibitemOpen
  \bibfield  {author} {\bibinfo {author} {\bibfnamefont {P.~P.}\ \bibnamefont
  {Ewald}},\ }\bibfield  {title} {\enquote {\bibinfo {title} {Die {Berechnung}
  optischer und elektrostatischer {Gitterpotentiale} (the calculation of
  optical and electrostatic grid potential)},}\ }\href {\doibase
  10.1002/andp.19213690304} {\bibfield  {journal} {\bibinfo  {journal} {Ann.
  Phys (Leipzig)}\ }\textbf {\bibinfo {volume} {369}},\ \bibinfo {pages}
  {253--287} (\bibinfo {year} {1921})}\BibitemShut {NoStop}%
\bibitem [{\citenamefont {Demyanov}\ and\ \citenamefont
  {Levashov}(2022)}]{DemLev22}%
  \BibitemOpen
  \bibfield  {author} {\bibinfo {author} {\bibfnamefont {G.~S.}\ \bibnamefont
  {Demyanov}}\ and\ \bibinfo {author} {\bibfnamefont {P.~R.}\ \bibnamefont
  {Levashov}},\ }\bibfield  {title} {\enquote {\bibinfo {title} {Systematic
  derivation of angular-averaged {Ewald} potential},}\ }\href {\doibase
  10.1088/1751-8121/ac870b} {\bibfield  {journal} {\bibinfo  {journal} {J.
  Phys. A: Math. Theor.}\ }\textbf {\bibinfo {volume} {55}},\ \bibinfo {pages}
  {385202} (\bibinfo {year} {2022})}\BibitemShut {NoStop}%
\bibitem [{\citenamefont {Lucha}\ \emph {et~al.}(1992)\citenamefont {Lucha},
  \citenamefont {Rupprecht},\ and\ \citenamefont
  {Schoeberl}}]{LuchaRuppScho992}%
  \BibitemOpen
  \bibfield  {author} {\bibinfo {author} {\bibfnamefont {W.}~\bibnamefont
  {Lucha}}, \bibinfo {author} {\bibfnamefont {H.}~\bibnamefont {Rupprecht}}, \
  and\ \bibinfo {author} {\bibfnamefont {F.~F.}\ \bibnamefont {Schoeberl}},\
  }\bibfield  {title} {\enquote {\bibinfo {title} {Significance of relativistic
  wave equations for bound states},}\ }\href {\doibase
  10.1103/PhysRevD.46.1088} {\bibfield  {journal} {\bibinfo  {journal} {Phys.
  Rev. D}\ }\textbf {\bibinfo {volume} {46}},\ \bibinfo {pages} {1088--1095}
  (\bibinfo {year} {1992})}\BibitemShut {NoStop}%
\bibitem [{\citenamefont {Plamondon}(2018)}]{Plam18}%
  \BibitemOpen
  \bibfield  {author} {\bibinfo {author} {\bibfnamefont {R.}~\bibnamefont
  {Plamondon}},\ }\bibfield  {title} {\enquote {\bibinfo {title} {General
  relativity: {An} erfc metric},}\ }\href {\doibase
  doi:10.1016/j.rinp.2018.02.035} {\bibfield  {journal} {\bibinfo  {journal}
  {Results in Physics}\ }\textbf {\bibinfo {volume} {9}},\ \bibinfo {pages}
  {456–462} (\bibinfo {year} {2018})}\BibitemShut {NoStop}%
\bibitem [{\citenamefont {Kestner}\ and\ \citenamefont
  {Sinanoglu}(1962)}]{KestSina62}%
  \BibitemOpen
  \bibfield  {author} {\bibinfo {author} {\bibfnamefont {N.~R.}\ \bibnamefont
  {Kestner}}\ and\ \bibinfo {author} {\bibfnamefont {O.}~\bibnamefont
  {Sinanoglu}},\ }\bibfield  {title} {\enquote {\bibinfo {title} {Study of
  electron correlation in helium-like systems using an exactly soluble
  model},}\ }\href {\doibase 10.1103/PhysRev.128.2687} {\bibfield  {journal}
  {\bibinfo  {journal} {Phys. Rev.}\ }\textbf {\bibinfo {volume} {128}},\
  \bibinfo {pages} {2687} (\bibinfo {year} {1962})}\BibitemShut {NoStop}%
\bibitem [{\citenamefont {Taut}(1993)}]{Taut93}%
  \BibitemOpen
  \bibfield  {author} {\bibinfo {author} {\bibfnamefont {M.}~\bibnamefont
  {Taut}},\ }\bibfield  {title} {\enquote {\bibinfo {title} {Two electrons in
  an external oscillator potential: {Particular} analytic solutions of a
  {Coulomb} correlation problem},}\ }\href {\doibase 10.1103/PhysRevA.48.3561}
  {\bibfield  {journal} {\bibinfo  {journal} {Phys. Rev. A}\ }\textbf {\bibinfo
  {volume} {48}},\ \bibinfo {pages} {3561--3566} (\bibinfo {year}
  {1993})}\BibitemShut {NoStop}%
\bibitem [{\citenamefont {Mandal}\ \emph {et~al.}(2003)\citenamefont {Mandal},
  \citenamefont {Mukherjee},\ and\ \citenamefont {Diercksen}}]{ManMukDie03}%
  \BibitemOpen
  \bibfield  {author} {\bibinfo {author} {\bibfnamefont {S.}~\bibnamefont
  {Mandal}}, \bibinfo {author} {\bibfnamefont {P.~K.}\ \bibnamefont
  {Mukherjee}}, \ and\ \bibinfo {author} {\bibfnamefont {G.~H.~F.}\
  \bibnamefont {Diercksen}},\ }\bibfield  {title} {\enquote {\bibinfo {title}
  {Two electrons in a harmonic potential: an approximate analytical
  solution},}\ }\href {\doibase doi:10.1088/0953-4075/36/22/009} {\bibfield
  {journal} {\bibinfo  {journal} {J. Phys. B: At. Mol. Opt. Phys.}\ }\textbf
  {\bibinfo {volume} {36}},\ \bibinfo {pages} {4483--4494} (\bibinfo {year}
  {2003})}\BibitemShut {NoStop}%
\bibitem [{\citenamefont {Karwowski}\ and\ \citenamefont
  {Cyrnek}(2004)}]{KarCyr03}%
  \BibitemOpen
  \bibfield  {author} {\bibinfo {author} {\bibfnamefont {J.}~\bibnamefont
  {Karwowski}}\ and\ \bibinfo {author} {\bibfnamefont {L.}~\bibnamefont
  {Cyrnek}},\ }\bibfield  {title} {\enquote {\bibinfo {title} {Harmonium},}\
  }\href {\doibase 10.1002/andp.200310071} {\bibfield  {journal} {\bibinfo
  {journal} {Ann. Phys. (Leipzig)}\ }\textbf {\bibinfo {volume} {13}},\
  \bibinfo {pages} {181 – 193} (\bibinfo {year} {2004})}\BibitemShut
  {NoStop}%
\end{thebibliography}%

\end{document}